\documentclass[12pt]{iopart}

\usepackage{graphicx}%
\usepackage{epsfig}
\usepackage{color}  
\begin{document}

\title[High energy photoelectron emission from gases using plasmonics enhanced near-fields]{High energy photoelectron emission from gases using plasmonics enhanced near-fields}

\author{M F Ciappina$^{1,2*}$, T Shaaran$^{1,3}$, R Guichard$^{4}$, J A P\'erez-Hern\'andez$^{5}$, L Roso$^{5}$, M Arnold$^{6}$,
T Siegel$^{6}$, A Za{\"i}r$^{6}$ and M Lewenstein$^{1,7}$}
\address{$^1$ICFO - Institut de Ci\`{e}nces Fot\`{o}niques, 08860 Castelldefels (Barcelona), Spain}
\address{$^2$Department of Physics, Auburn University, Auburn, Alabama 36849, USA}
\address{$^3$CEA-Saclay, IRAMIS, Service des Photons, Atomes et Molecules, 91191
Gift-sur-Yvette, France}
\address{$^4$Laboratoire Chimie-Physique, Mati\`ere et Rayonnement, Universit\`e Pierre et Marie Curie, UMR 7614, F-75231 Paris Cedex 5, France}
\address{$^{5}$Centro de L\'aseres Pulsados, Parque Cient\'{\i}fico, E-37185 Villamayor, Salamanca, Spain}
\address{$^{6}$Blackett Laboratory Laser Consortium, Department of Physics, Imperial College London, London SW7 2AZ, United Kingdom}
\address{$^7$ICREA-Instituci\'o Catalana de Recerca i Estudis Avan\c{c}ats, Lluis Companys 23, 08010 Barcelona, Spain}
\ead{$^{*}$mfc0007@auburn.edu}
\begin{abstract}
We study theoretically the photoelectron emission in noble gases using plasmonic enhanced near-fields. We demonstrate that these fields have a great potential to generate high energy electrons by direct mid-infrared laser pulses of the current femtosecond oscillator. Typically, these fields appear in the surroundings of plasmonic nanostructures, having different geometrical shape such as bow-ties, metallic waveguides, metal nanoparticles and nanotips, when illuminated by a short laser pulse. In here, we consider metal nanospheres, in which the spatial decay of the near-field of the isolated nanoparticle can be approximated by an exponential function according to recent attosecond streaking measurements. We establish that the strong nonhomogeneous character of the enhanced near-field plays an important role in the above threshold ionization (ATI) process and leads to a significant extension in the photoelectron spectra. In this work, we employ the time dependent Schr\"odinger equation in reduced dimensions to calculate the photoelectron emission of xenon atoms in such enhanced near-field. Our findings are supported by classical calculations.

\end{abstract}

\pacs{42.65.Ky,78.67.Bf, 32.80.Rm}
\submitto{Las. Phys. Lett.}
\maketitle

Above-threshold ionization (ATI) has been a particular and interesting subject in both experimental and theoretical physics. The ATI phenomenon, which was experimentally observed more than three decades ago~\cite{Agostini1979}, occurs when an atom or molecule absorbs more photons than the minimum number required to be ionized, the remaining energy is then converted into kinetic energy for the released electron.

Recent advances in the laser technology allow to routinely generate few-cycle pulses, which found a wide range of applications in science, such as controlling chemical reactions and molecular motion in their natural scales~\cite{schnurer2000,vdHoff2009}, generating high order harmonic and even creating isolated extreme ultraviolet (XUV) pulses~\cite{ferrari2010,schultze2007}, which enable us even more control in a sub-femtosecond and sub-{\AA}ngstr\"om temporal and spatial scales, respectively.

In a few-cycle pulse the laser electric field can be fully characterized by its duration and by the so-called carrier-envelope phase (CEP). In contrast to multicycle pulses, the electric field of a few-cycle pulse, and consequently the laser-matter processes driven by it, are greatly affected by the CEP~\cite{Wittmann2009,kling2008}. The importance of CEP, has been experimentally observed in high-harmonic generation (HHG)~\cite{nisoli2003}, the direct emission of electrons from atoms~\cite{paulus2001}, and in the nonsequential double ionization yield~\cite{liu2004}. In order to have a better control of the system in a sub-temporal temporal scale, it is important to find reliable schemes to measure and precisely characterize the absolute phase of few-cycle pulses.

The sensitivity of the energy and angle-resolved photoelectron spectra to the absolute value of the CEP has sparkled the investigation of ATI generated by few-cycle driving laser pulses. Consequently, this feature configures the ATI phenomenon as a very reliable and robust tool in laser pulse characterization. In order to extract the CEP of a few-cycle laser pulse, the backward-forward asymmetry of the ATI spectrum has to be measured and from the information collected, the absolute CEP can be obtained~\cite{paulus2003}. It is also established that the absolute CEP can be measured more accurately if the ATI photoelectron kinetic energy spectrum has a larger cutoff.

New experiments in noble gases have demonstrated that the harmonic cutoff could be extended by using plasmon field enhancement~\cite{kim,kling}. This field appears when a nanostructure is illuminated by a short laser pulse. As a result, the external femtosecond low intensity pulse couples with the plasmon mode inducing a collective oscillation of free charges within the localized regions of the nanostructure and forming a spot of highly enhanced electric field. Due to the strong confinement of the plasmonic spots and the distortion of the electric field by surface plasmons, the locally enhanced field is not spatially homogeneous and strongly influences the subsequent motion of the electron in the continuum. Consequently, important changes will occur to the main features of the strong field phenomena~\cite{kim,kling}, since now the force applied to the active electron depends also on position. Strong field phenomena in such kind of fields has been a topic of intense activity nowadays~\cite{husakou,ciappi2012,yavuz1,ciappi_opt,ciappiati,tahirsfa,joseprl,tahirrapid,yavuz2}.

Recently, in the context of strong fields, an alternative process has been employed to produce photoelectron emission using metallic solid state nanostructures targets instead of atoms and molecules in gas phase. This phenomenon, which is known as Above Threshold Photoemission (ATP), has received special attention due to its novelty and the new physics involved. Due the plasmonic field enhancement, the electrons emitted by ATP from metallic surfaces or metal nanotips present distinct characteristics, namely higher kinetic energies in a non-ponderomotive regime~\cite{ropers}. In comparison to the noble gases interacting with the same strong laser field, in here the photoelectron kinetic energy spectrum has much larger cutoff (see e.g.~\cite{ropers,peterprl2006,peterprl2010,peternature,peterjpbreview}).

In this Letter we put forward the plausibility to perform ATI experiments by combining plasmonic enhanced near-fields and noble gases. The proposed experiment would take advantage of the plasmonic enhanced near-fields, which present a strong spatial nonhomogeneous character and the flexibility to use any atom or molecule in gas phase. From the theoretical viewpoint the ATI process can be tackled using different approaches (for a summary see e.g.~\cite{milosevic_rev,schafer1993,telnov2009,bauer2006,Blaga2009,Quan2009} and references therein). We employed the numerical solution of the Time Dependent Schr\"odinger Equation (TDSE) in reduced dimensions by including the actual functional form of metal nanoparticles plasmonic near-fields obtained from attosecond streaking measurements~\cite{kling_spie,kling_prb}. We have chosen this particular nanostructure since its actual enhanced-field is known experimentally, while for the other nanostructures like bow-ties~\cite{kim} the real plasmonic field is unknown. For the most of the plasmonic nanostructures the enhanced field is theoretically calculated using the finite element simulation, which is based on an ideal system, not the complex situation we are dealing with. For instance, Ref.~\cite{kim} states an intensity enhancement of 4 order of magnitude (calculated theoretically) but the maximum harmonic order measured was 17, which corresponds to the intensity enhancement of 2 order of magnitude rather than 4 (for more details see~\cite{ciappi_opt,tahirjmo}). However, our numerical tool is developed in such a way to allow the treatment of a very general set of non-homogeneous fields such as those present in the vicinity of metal nanostructures~\cite{kim}, dielectric nanoparticles~\cite{kling2008,kling}, or metal nanotips~\cite{ropers}. The kinetic energy for the electrons both direct and rescattered are classically calculated and compared to our quantum mechanical approach predictions.

The one-dimensional time-dependent Schr\"odinger equation (1D-TDSE) for a model atom reads:
\begin{equation}
\label{tdse}
\rmi \frac{\partial \Psi(x,t)}{\partial t}=\mathcal{H}(t)\Psi(x,t)=\left[-\frac{1}{2}\frac{\partial^{2}}{\partial x^{2}}+V_{a}(x)+V_{l}(x,t)\right]\Psi(x,t)
\end{equation}
where $V_{l}(x,t)$ is the laser-atom interaction. For the atomic
$V_{a}(x)$ potential, we use the quasi-Coulomb or soft core potential $V_{a}(x)=-\frac{1}{\sqrt{x^2+a^2}}$,
which was firstly introduced in~\cite{eberly} and has been widely used in the study
of laser-matter processes in atoms. The parameter $a$ allows us to tune the ionization potential of the atom under consideration.
In our studies we consider the field to be linearly polarized along the $x$-axis and modify
the interaction term $V_{l}(x,t)$ in order to treat spatially
non-homogeneous near-fields but still maintaining the dipole character.
Consequently we write
\begin{eqnarray}  \label{vlaser}
V_{l}(x,t)&=&-E(x,t)\,x
\end{eqnarray}
where $E(x,t)$ is the laser electric field characterized by
\begin{equation}  \label{electric}
E(x,t)=E_0(t)\exp(-x/\chi).
\end{equation}
In equation~(\ref{electric}), we have collected all the temporal information in $E_0(t)$, i.e. $E_0(t)=E_p f(t) \sin(\omega t+\phi)$. In here, $E_p$, $f(t)$, $\omega$ and $\phi$ are the peak amplitude of the laser electric field, the pulse envelope, the frequency of the laser pulse and the
CEP, respectively. The exponential part of equation~(\ref{electric}) corresponds to the spatial dependency of the plasmonic enhanced field, where $\chi$
is a parameter that characterizes the spatial decay of the plasmonic near-field and depends on both the geometrical and material characteristics~\cite{ciappi_opt,kling} (note the units of $\chi$ are in length).
The equation~(\ref{electric}) is only valid for $x\ge R_0$, where $R_0$ is the radius of the metal nanoparticle, i.e. for values of $x$ that is outside of the metal nanoparticle surface. In addition, the electron motion is in the region $x\ge R_0$ with $(x+R_0)\gg0$.
To model short laser pulses, we use a sin$^2$ envelope $f(t)$ of the form $f(t)=\sin^{2}\left(\frac{\omega t}{2 n_p}\right)$, where $n_p$ is the total number of optical cycles. The time duration of the laser pulse will be $T_p=n_p \tau$, where $\tau=2\pi/\omega$ denotes the laser period. We assume that the atomic target is in the ground state ($1s$) before we turn on the
laser ($t=-\infty$). This particular state can be found by solving an eigenvector and eigenvalue problem once the 1D-TDSE has been discretized. For Xe atoms we chose $a=1.62$ to match its atomic ionization potential $I_p=12.199$ eV (0.446 a.u.).

The equation~(\ref{tdse}) was solved numerically by
using the Crank-Nicolson scheme with an adequate spatial grid~\cite{keitel}. In our numerical experiment we employed boundary reflection mask functions~\cite{mask} in order to avoid spurious contributions and the grid size has been adequately extended in order to treat high energy electrons. It is well known these electrons can travel far away from their parent nucleus without returning to the core, however, this effect was considered in our numerical scheme. For calculating the energy-resolved photoelectron spectra $P(E)$ we used the
window function technique developed by Schafer~\cite{schaferwop1,schaferwop}. This tool represents an improvement with respect to the usual projection methods and has been widely used for calculating angle-resolved and
energy-resolved photoelectron spectra~\cite{schaferwop2}. 

In Fig.~1, we presented a schematic picture of the systems under study in which a metal nanoparticle as those used in the experiment of Ref~\cite{kling_spie}, is surrounded by Xe atoms. We also included the time and spatial profile of the laser electric field where it is possible to observe the usual time dependence and the strong spatial variation at a nanometer scale.
In this nanoparticle the spatial variation of the electric field produces strong modifications in the electron motion and consequently in the associated laser-matter processes, for example above-threshold ionization (ATI). Furthermore each metal nanoparticle would acts as a laser nanosource which contributes to photoelectron emission process. 

In our model we have employed the laser field enhanced parameters used in the experiment of Ref.~\cite{kling} ($I=2\times10^{13}$ W/cm$^{2}$ and $\lambda=720$ nm) to make our proposed experiment perfectly feasible. For this kind of metallic nanotip, the enhanced laser field intensity corresponds to initial laser input intensities of $10^{12}$ W/cm$^{2}$, which through plasmon enhanced amplification process is intensified by more than one order of magnitude (see~\cite{peterjpbreview} and references therein). The gaussian shaped experimental laser pulses with 5 fs FWHM are modeled by employing a sin$^{2}$ shaped pulse with 5 optical cycles (13 fs). 

As a first test we maintain the aforementioned laser field enhanced parameters and we varied the $\chi$ parameter in equation~(\ref{electric}). In Fig.~2, we present the photoelectron spectra calculated using 1D-TDSE for Xe atoms. Each curve present different values of $\chi$: homogeneous case ($\chi\rightarrow\infty$), $\chi=40$, $\chi=35$ and $\chi=29$. For the homogeneous case there is a visible cutoff at $\approx 10.5$ eV confirming the well known ATI cutoff at $10U_{p}$, which corresponds to those electrons that once ionized return to the core and elastically rescatter. In here $U_{p}$ is the ponderomotive potential given by $U_{p}=E_{p}^{2}/4\omega^{2}$.  This value in perfect agreement with the one shown in Ref.~\cite{kling}. On the other hand, for this particular intensity the cutoff at $2U_{p}$ ($\approx 2.1$ eV) developed by the direct ionized electrons is not visible in the spectra. For the nonhomogeneous cases the cutoff of the rescattered electron is far beyond the classical limit $10U_{p}$, depending on the $\chi$ parameter chosen. As it is depicted in Fig.~2 the cutoff is extended as we decrease the value of $\chi$. For $\chi= 40$ the cutoff is at around $14$ eV, while for $\chi= 29$ it is around $30$ eV. The low energy region of the photoelectron spectra is sensitive to the atomic potential of the target and one needs to calculate TDSE in full dimensionality in order to model this region adequately. In this paper we are interested in the high energy region of the photoelectron spectra, which  is very convenient because it isn't greatly affected by the considered atom. Thus by employing 1D-TDSE the conclusions that can be taken from these high energetic electrons are very reliable.
 
In order to investigate fully the behavior of the system, i.e. the metallic nanoparticles surrounded with noble gas atoms, we considered a larger laser field intensity of $I=5\times10^{13}$ W/cm$^{2}$ by keeping the rest of the laser parameters the same. In Fig.~3, we have plotted the photoelectron spectra for the homogeneous case and for $\chi=29$. From this plot we observe that the nonhomogeneous character of the laser enhanced electric field introduces a highly nonlinear behavior in this particular laser-matter phenomenon. For this intensity with $\chi=29$ it is possible to obtain very energetic electrons reaching values of several hundreds of eV. This is a good indication that the nonlinear behavior of the combined system of the metallic nanoparticles and noble gas atoms could pave the way to generate keV electrons with tabletop laser sources.

We now concentrate our efforts in order to understand the extension of the
energy-resolved photoelectron spectra using classical arguments. From the
simple-man model~\cite{corkum} we can describe the physical origin of the
ATI process as follows: an electron bound to its parent ion at a position $x=0$, is liberated
to the continuum at a given time called the \textit{ionization} or born time $t_{i}$, but
with zero velocity, i.e. $\dot{x}(t_{i})=0$. In the strong-field picture this electron now experiences
only the influence of the oscillating laser electric field (the
residual Coulomb interaction is neglected) and will reach
the detector either directly or through the so-called \textit{rescattering} process. By using
the Newton equations it is possible to calculate the maximum
kinetic energy of the electron for both the direct and rescattered processes. The classical
equation of motion for an electron in the laser field can be written as $\ddot{x}(t) =-\nabla _{x}V_{l}(x,t)$. More specifically, for the electric field of the form (\ref{electric}) it reads $\ddot{x}(t) =-E(x,t)(1-x(t)/\chi)$. Note that in the limit $\chi\rightarrow\infty$, we recover the conventional homogeneous case $\ddot{x}(t) =-E(t)$. 

For the direct ionization the electron reaches the detector with a kinetic energy given by
\begin{equation}
\label{direct}
E_{d}=\frac{\left[ \dot{x}(t_{i})-\dot{x}(t_{f})\right]^{2}}{2},
\end{equation}
where $t_{f}$ is the end time of the laser pulse. For the rescattered
ionization, in which the electron returns to the core at a time $t_{r}$ and
reverses its direction, the kinetic energy  of the electron yields
\begin{equation}
\label{rescattered}
E_{r}=\frac{\left[ \dot{x}(t_{i})+\dot{x}(t_{f})-2\dot{x}(t_{r})\right]^{2}}{2}.
\end{equation}

For homogeneous fields equations~(\ref{direct}) and (\ref{rescattered}) result $E_{d}=\frac{\left[ A(t_{i})-A(t_{f})\right] ^{2}}{2}$ and
$E_{r}=\frac{\left[ A(t_{i})+A(t_{f})-2A(t_{r})\right] ^{2}}{2}$, with $A(t)$ being the laser vector potential $A(t)=-\int^{t} E(t')dt'$. Furthermore, it is well known that the maximum values for $E_{d}$ and $E_{r}$ are $2U_{p}$ and $10U_{p}$, respectively~\cite{milosevic_rev}. Although for laser electric fields of the form given by equation~(\ref{electric}), the aforementioned limits should to be revised.

In Fig.~4 we present the numerical solutions of the Newton equation for an electron moving in a linearly polarized spatially nonhomogeneous electric field of the form (\ref{electric}), in terms of the kinetic energy of the direct and rescattered electrons. We employed the same laser enhanced field parameters as in Figs.~2 and 3. Panels
(a) and (c) show the kinetic energy of the direct ($\color{green}\bullet $) and rescattered ($\color{red}\bullet$) electrons for the homogeneous case with laser enhanced intensities of $I=2\times10^{13}$ W/cm$^{2}$ and $I=5\times10^{13}$ W/cm$^{2}$, respectively. In panels (b) and (d), we plotted the kinetic energy of the direct ($\color{green}\bullet$) and rescattered ($\color{red}\bullet$) electrons for the extreme nonhomogeneous case ($\chi=29$) and for the mentioned intensities. Furthermore, we observed an unconventional behavior due to the nonhomogeneous character of the enhanced near-fields. From the panels (b) and (d), we clearly see that spatially nonhomogeneous enhanced fields produce electrons with high kinetic energy, well far beyond the usual classical limits of $2U_p$ and $10U_p$ for the direct and rescattered process respectively. These new features are related to the changes in the electron trajectories (for details see e.g.~\cite{yavuz1,ciappi2012,ciappi_opt}). In here the electron trajectories are
substantially modified in such a way that the released electron has longer time to spend in the continuum
acquiring more energy from the laser electric field. For the homogeneous case on the other hand classical simulations show both the $2U_p$ (direct electrons) and $10U_p$ (rescattered electrons) kinetic energy limits and the values are in perfect agreement with the 1D-TDSE simulations. A similar behavior was observed in metal nanotips~\cite{ropers}, although in our case both direct and rescattered electrons are considered.

In this Letter we propose generation of high energy photoelectrons using near-enhanced fields by combining metallic nanoparticles and noble gas atoms. Near-enhanced fields present a strong spatial dependence at a nanometer scale and this behavior introduces substantial changes in the laser-matter processes. We have modified the time dependent Schr\"odinger equation to model the ATI phenomenon in noble gases driven by the enhanced near-fields of such nanostructure. We predict a substantial extension in the cutoff position of the energy-resolved photoelectron spectra, far beyond the conventional $10U_p$ classical limit. These new features are well reproduced by classical simulations. Our predictions would pave the way to the production of high energy photoelectrons reaching the keV regime by using a combination of metal nanoparticles and noble gases. In this kind of system each metal nanoparticle configures a laser nanosource with particular characteristics that allow not only the amplification of the input laser field, but also the modification of the laser-matter phenomena due to the strong spatial dependence of the generated coherent electromagnetic radiation.

We acknowledge the financial support of the MICINN projects (FIS2008-00784
TOQATA, FIS2008-06368-C02-01 and FIS2010-12834); ERC Advanced Grant
QUAGATUA, Alexander von Humboldt Foundation and Hamburg Theory Prize (M.
L.). This research has been partially supported by Fundaci\'o Privada Cellex. J. A. P.-H. acknowledges support from Spanish
MINECO through the Consolider Program SAUUL
(CSD2007-00013) and research project FIS2009-09522,
from Junta de Castilla y Le\'on through the Program for
Groups of Excellence (GR27) and from the ERC Seventh
Framework Programme (LASERLAB-EUROPE, Grant
No. 228334). A. Z. acknowledges the support from EPSRC Grant No. EP/J002348/1 and Royal Society International Exchange Scheme 2012 Grant No. IE120539.

\newpage
Figure Captions

\begin{figure}[htb]
\centering
\includegraphics[width=0.75\textwidth]{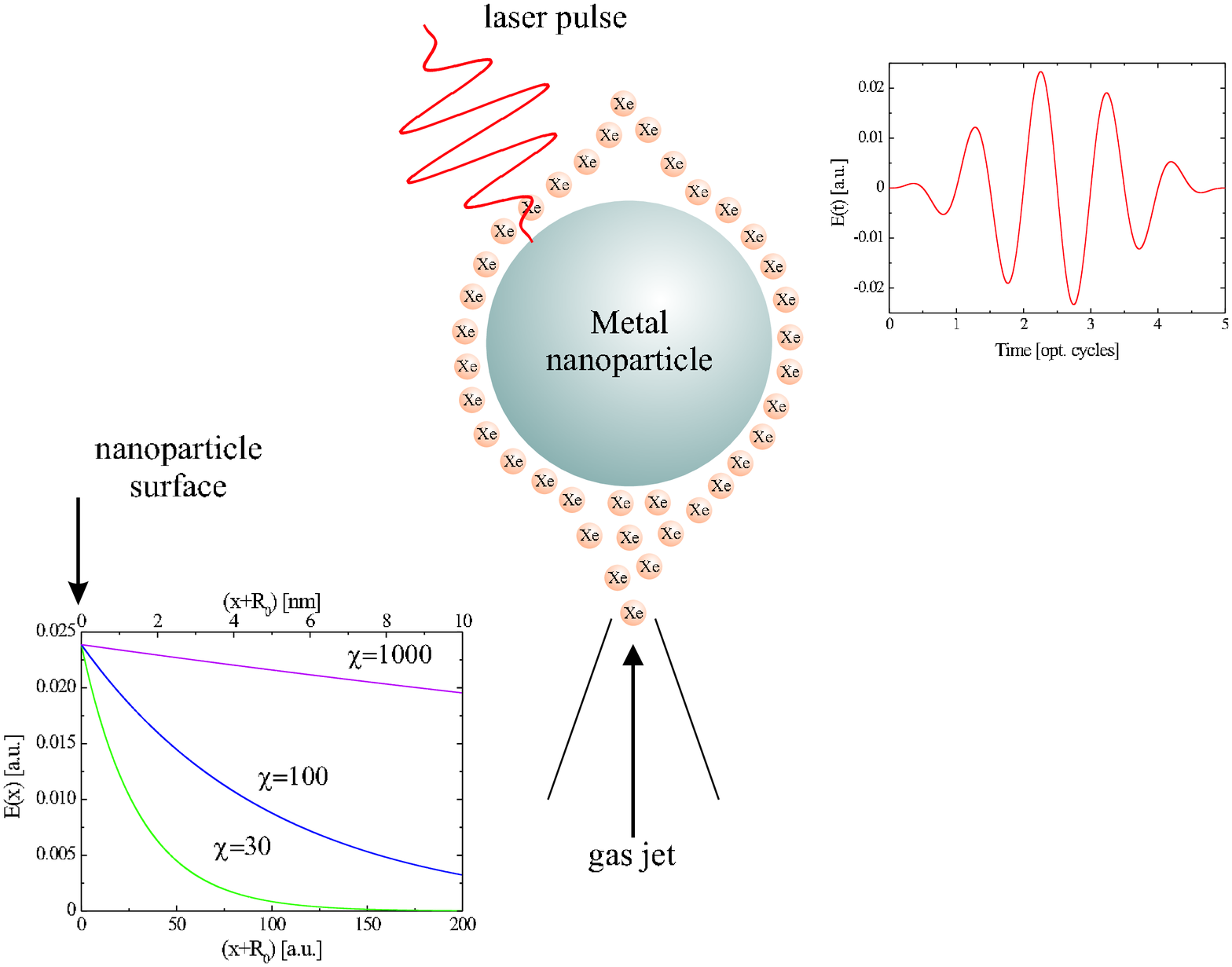}
\caption{(color online) Schematic picture of the proposed system to generate high energy electrons. The upper right and lower left insets show the temporal and spatial dependence of the enhanced near-field, respectively.}
\label{fig:figure1}
\end{figure}

\newpage

\newpage
\begin{figure}[htb]
\centering
\vspace{2cm}
\hspace{-2.5cm}
\includegraphics[width=0.6\textwidth]{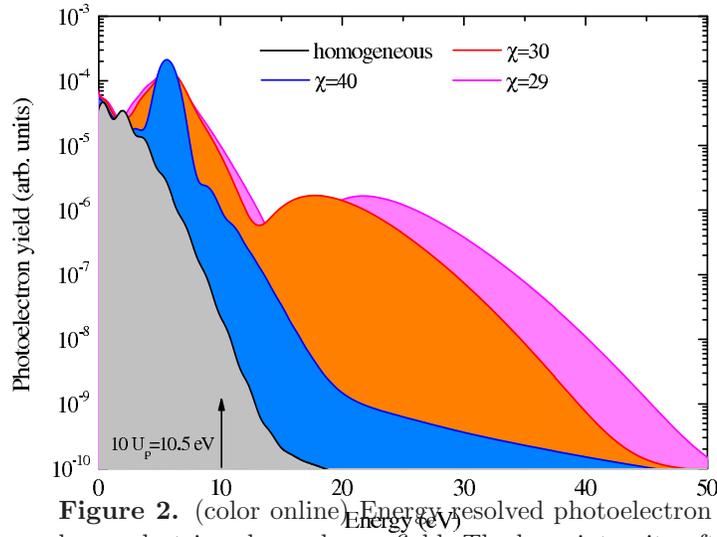}
\vspace{-1cm}
\caption{(color online) Energy resolved photoelectron spectra for Xe atoms driven by an electric enhanced near-field. The laser intensity after interacting with the metal nanoparticles is $I=2\times10^{13}$ W/cm$^{2}$. We employ $\phi=\pi/2$ (cos-like pulses) and the laser wavelength and number of cycles remain unchanged with respect to the input pulse, i.e. $\lambda=720$ nm and $n_p=5$ (13 fs in total). The arrow indicates the conventional $10 U_p$ cutoff (10.5 eV for this case).}
\label{fig:figure2}
\end{figure}

\newpage

\begin{figure}[htb]
\centering
\vspace{2cm}
\hspace{-2.5cm}
\includegraphics[width=0.6\textwidth]{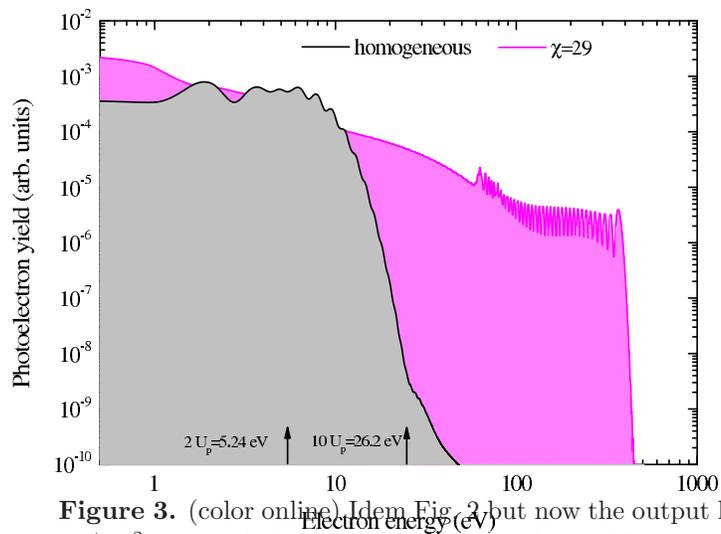}
\vspace{-1cm}
\caption{(color online) Idem Fig.~2 but now the output laser intensity is $I=5\times10^{13}$ W/cm$^{2}$. We include arrows for the two classical limits, $2 U_p$ (5.24 eV) and $10U_p$ (26.2 eV), respectively.}
\label{fig:figure3}
\end{figure}

\newpage

\begin{figure}[htb]
\centering
\includegraphics[width=0.75\textwidth]{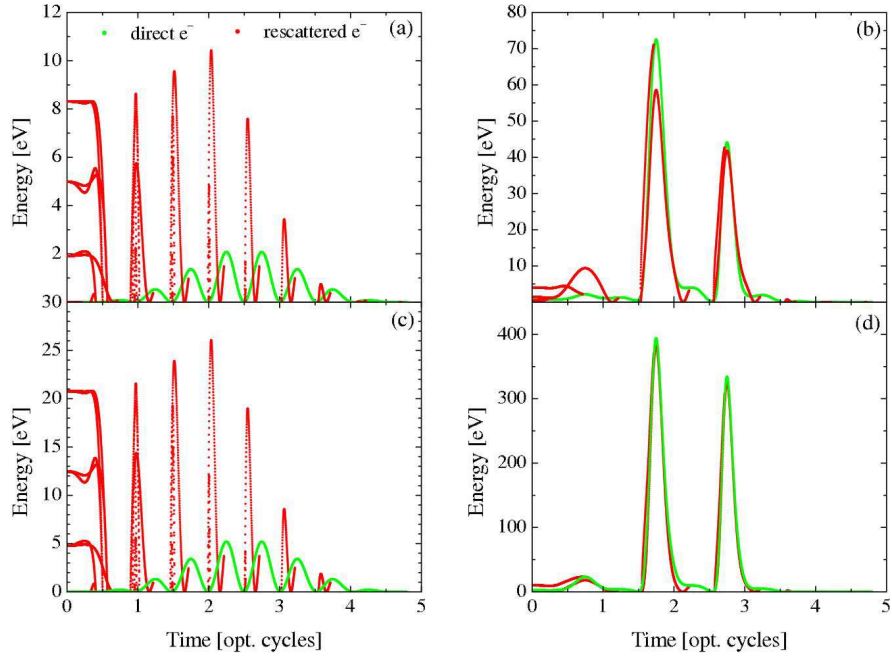}
\caption{(color online) Kinetic energy of the direct ($\color{green}\bullet$) and rescattered ($\color{red}\bullet$) electrons obtained from the classical equations of motion. Panels (a) and (c) are the homogeneous case for laser intensities $I=2\times10^{13}$ W/cm$^{2}$ and $I=5\times10^{13}$ W/cm$^{2}$, respectively and panels (b) and (d) are the nonhomogeneous case with $\chi=29$ and for laser intensities $I=2\times10^{13}$ W/cm$^{2}$ and $I=5\times10^{13}$ W/cm$^{2}$, respectively.}
\label{fig:figure4}
\end{figure}

\end{document}